\documentstyle[aps,prl,multicol]{revtex}
\input epsf

\begin{document}

\title{GENERALIZED UNCERTAINTY PRINCIPLE AND DARK MATTER}

\author{Pisin Chen}

\address{Stanford Linear Accelerator Center \\ Stanford University, Stanford, CA 94309, USA
     }

\maketitle

\begin{abstract}
There have been proposals that primordial black hole remnants (BHRs) are the
dark matter, but the idea is somewhat vague. Recently we argued that the
generalized uncertainty principle (GUP) may prevent black holes from
evaporating completely, in a similar way that the standard uncertainty
principle prevents the hydrogen atom from collapsing. We further noted that
the hybrid inflation model provides a plausible mechanism for production of
large numbers of small black holes. Combining these we suggested that the
dark matter might be composed of Planck-size BHRs. In this paper we briefly
review these arguments, and discuss the reheating temperature as a result of
black hole evaporation.
\end{abstract}

\section{Introduction}

It is by now widely accepted that dark matter (DM) constitutes a substantial
fraction of the present critical energy density in the universe. However, the
nature of DM remains an open problem. There exist many DM candidates, among
which a contending category is weakly interacting massive particles, or
WIMPs. It has been suggested that primordial black holes
(PBHs)\cite{zeldovich,hawking71} are a natural candidate for
WIMPs\cite{macgibbon}. More recent studies\cite{carr94} based on the PBH
production from the ``blue spectrum" of inflation demand that the spectral
index $n\sim 1.3$, but this possibility may be ruled out by the recent WMAP
experiment\cite{WMAP}.

In the standard view of black hole thermodynamics, based on the entropy
expression of Bekenstein\cite{bekenstein1} and the temperature expression of
Hawking\cite{hawking}, a small black hole should emit blackbody radiation,
thereby becoming lighter and hotter, leading to an explosive end when the
mass approaches zero. However Hawking's calculation assumes a classical
background metric and ignores the radiation reaction, assumptions which must
break down as the black hole becomes very small and light. Thus it does not
provide an answer as to whether a small black hole should evaporate entirely,
or leave something else behind, which we refer to as a black hole remnant
(BHR).

Numerous calculations of black hole radiation properties have been made from
different points of view\cite{wilczek}, and some hint at the existence of
remnants, but none appears to give a definitive answer. A cogent argument
against the existence of BHRs can be made\cite{susskind}: since there is no
evident symmetry or quantum number preventing it, a black hole should radiate
entirely away to photons and other ordinary stable particles and vacuum, just
like any unstable quantum system.

In a series of recent papers\cite{acs,chen03}, a generalized uncertainty
principle (GUP)\cite{veneziano,adler,maggiore} was invoked to argue the
contrary, that the total collapse of a black hole may be prevented by
dynamics and not by symmetry, just like the prevention of hydrogen atom from
collapse by the uncertainty principle\cite{shankar}. These arguments then
lead to a modified black hole entropy and temperature, and as a consequence
the existence of a BHR at around the Planck mass. This notion was then
combined with hybrid inflation model\cite{linde91,copeland,randall,lyth99}
and it was shown that primordial BHRs might in principle be the primary
source for dark matte\cite{chen03}. In this paper we briefly reproduce these
arguments, and include additional discussion on the reheating temperature as
a result of black hole evaporation.

\section{Generalized Uncertainty Principle}

As a result of string theory\cite{veneziano} or general considerations of
quantum mechanics and gravity\cite{adler,maggiore}, the GUP gives the
position uncertainty as
\begin{equation}
\Delta x\geq \frac{\hbar}{\Delta p}+ l_p^2\frac{\Delta p}{\hbar} \ ,
\label{eq:B}
\end{equation}
where $l_p=(G\hbar/c^3)^{1/2}\approx 1.6\times 10^{-33}$cm is the Planck
length. A heuristic derivation may also be made on dimensional grounds. We
think of a particle such as an electron being observed by means of a photon
with momentum $p$. The usual Heisenberg argument leads to an electron
position uncertainty given by the first term in Eq.(\ref{eq:B}). But we
should add to this a term due to the gravitational interaction of the
electron with the photon, and that term must be proportional to $G$ times the
photon energy, or $Gpc$. Since the electron momentum uncertainty $\Delta p$
will be of order of $p$, we see that on dimensional grounds the extra term
must be of order $G\Delta p/c^3$, as given in Eq.(\ref{eq:B}). Note that
there is no $\hbar$ in the extra term when expressed in this way. The
position uncertainty has a minimum value of $\Delta x=2l_p$, so the Planck
distance, $l_p$, plays the role of a fundamental length.

\section{Black Hole Remnant}

The characteristic energy $E$ of the emitted photons may be estimated from
the uncertainty principle. In the vicinity of the black hole surface there is
an intrinsic uncertainty in the position of any particle of about the
Schwarzschild radius, $\Delta x\approx r_s$, due to the behavior of its field
lines\cite{adler2} - as well as on dimensional grounds. This leads to a
momentum uncertainty
\begin{equation}
\Delta p \approx \frac{\hbar}{\Delta
x}=\frac{\hbar}{r_s}=\frac{\hbar c^2}{2GM_{\rm BH}}\ ,
\label{eq:C}
\end{equation}
and hence to an energy uncertainty of $\Delta pc\approx \hbar c^3/2GM_{\rm
BH}$. We identify this as the characteristic energy of the emitted photon,
and thus as a characteristic temperature; it agrees with the Hawking
temperature up to a factor $4\pi$, which we will henceforth include as a
``calibration factor" and write (with $k_B=1$),
\begin{equation}
T_{\rm H} \approx \frac{\hbar c^3}{8\pi GM_{\rm BH}}=\frac{M_p^2 c^2}{8\pi
M_{\rm BH}} \ , \label{eq:D}
\end{equation}
where $M_p=(\hbar c/G)^{1/2}\approx 1.2\times 10^{19}$GeV is the Planck mass.

The blackbody energy output rate of BH is given by
\begin{equation}
\dot{x}=\frac{1}{t_{ch}(x_i^3-3t/t_{ch})^{2/3}}\ , \label{eq:F}
\end{equation}
where $x=M_{\rm BH}/M_p$ and $x_i$ refers to the initial mass of
the hole. $t_{ch}=60(16)^2\pi t_p\approx 4.8\times 10^4 t_p$ is a
characteristic time for BH evaporation, and $t_p=(\hbar
G/c^5)^{1/2}\approx 0.54\times 10^{-43}$sec is the Planck time.
The black hole thus evaporates to zero mass in a time given by
$t/t_{ch}=x_i^3/3$, and the rate of radiation has an infinite
spike at the end of the process.

The momentum uncertainty according to the GUP is
\begin{equation}
\frac{\Delta p}{\hbar}\approx\frac{\Delta x}{2l_p^2}\Big[1\mp
\sqrt{1-4l_p^2/(\Delta x)^2}\Big]\ . \label{eq:G}
\end{equation}
Therefore the modified black hole temperature becomes
\begin{equation}
T_{\rm GUP} =\frac{M_pc^2}{4\pi}x\Big[1\mp\sqrt{1-1/x^2}\Big]\ .
\label{eq:H}
\end{equation}
This agrees with the Hawking result for large mass if the negative sign is
chosen, whereas the positive sign has no evident physical meaning. Note that
the temperature becomes complex and unphysical for mass less than the Planck
mass and Schwarzschild radius less than $2l_p$. At the Planck mass the slope
is infinite, which corresponds to zero heat capacity of the black hole, and
the evaporation comes to a stop.

If there are $g$ species of relativistic particles, then the BH evaporation
rate is
\begin{equation}
\dot{x}=-\frac{16g}{t_{ch}}x^6\Big[1-\sqrt{1-1/x^2} \Big]^4 \ . \label{eq:I}
\end{equation}
Thus the hole with an initial mass $x_i$ evaporates to a Planck mass remnant
in a time given by
\begin{eqnarray}
\tau&=&\frac{t_{ch}}{16g}\Big[\frac{8}{3}x_i^3-8x_i-\frac{1}
{x_i}+\frac{8}{3}(x_i^2-1)^{3/2}-4\sqrt{x_i^2-1}+4\cos^{-1}\frac{1}{x_i}
+\frac{19}{3}\Big] \cr &\approx& \frac{x_i^3}{3g}t_{ch}, \quad\quad\quad
x_i\gg 1\ . \label{eq:J}
\end{eqnarray}
The energy output given by Eq.(\ref{eq:I}) is finite at the end point where
$x=1$, i.e., $dx/dt\vert_{x=1}=-16g/t_{ch}$, whereas for the Hawking case it
is infinite at the endpoint where $x=0$. The present result thus appears to
be more physically reasonable. The evaporation time in the $x_i\gg 1$ limit
agrees with the standard Hawking picture.

\section{Hybrid Inflation and Black Hole Production}

The hybrid inflation, first proposed by A. Linde\cite{linde91},
can naturally induce large number of small PBHs\cite{bellido}. In
the hybrid inflation model two inflaton fields, $(\phi,\psi)$, are
invoked. Governed by the inflation potential,
$\phi$ first executes a ``slow-roll" down the potential, and is
responsible for the more than 60 e-folds expansion while $\psi$
remains zero. When $\phi$ eventually reduces to a critical value,
it triggers a phase
transition that results in a ``rapid-fall" of the energy density
of the $\psi$ field, which lasts only for a few e-folds, that ends
the inflation.

The evolution of the $\psi$ field during the second stage
inflation, measured backward from the end, is
\begin{equation}
 \psi(N[t])= \psi_e\exp(-sN[t]) \ , \label{eq:N}
\end{equation}
where $N(t)=H_*(t_e-t)$ is the number of e-folds from $t$ to
$t_e$, $H_*$ is the Hubble parameter during inflation, and $s$ is
a numerical factor of the order unity.

Quantum fluctuations of $\psi$ induce variations of the starting
time of the second stage inflation, i.e., $\delta t =
\delta\psi/\dot{\psi}$. This translates into perturbations on the
number of e-folds, $\delta N=H_*\delta\psi/\dot{\psi}$, and
therefore the curvature contrasts, $\delta\rho/\rho\equiv \delta$.
With an initial density contrast $\delta(m)\equiv
\delta\rho/\rho\vert_m$, the probability that a region of mass $m$
becomes a PBH is\cite{carr75}
\begin{equation}
P(m) \sim \delta(m)e^{-w^2/2\delta^2} \ . \label{eq:Q}
\end{equation}

Let us assume that the universe had inflated $e^{N_c}$ times
during the second stage of inflation. It can be
shown\cite{bellido} that
\begin{equation}
e^{N_c}\sim \Big(\frac{2M_p}{sH_*}\Big)^{1/s}\ , \label{eq:R}
\end{equation}
and the curvature perturbations reentered the horizon at time
\begin{equation}
t\sim t_h=H_*^{-1}e^{3N_c} \ . \label{eq:S}
\end{equation}
At this time if the density contrast was $\delta \sim 1$, then BHs with size
$r_s\sim H_*^{-1}e^{3N_c}$ would form with an initial mass
\begin{equation}
M_{{\rm BH}i}\simeq \frac{M_p^2}{H_*}e^{3N_c} \ . \label{eq:T}
\end{equation}

Following the numerical example given in Ref.21, we let $H_*\sim 5\times
10^{13}$ GeV and $s\sim 3$. Then the density contrast can be shown to be
$\delta \sim 1/7$, and the fraction of matter in the BH is thus $P(m)\sim
10^{-2}$. From Eq.(\ref{eq:R}), $e^{N_c}\sim 54$. So the total number of
e-folds is $N_c\sim 4$. The black holes were produced at the moment $t_h\sim
2\times 10^{-33}$ sec, and had a typical mass $M_{{\rm BH}i}\sim 4\times
10^{10}M_p$. Let $g\sim 100$. Then the time it took for the BHs to reduce to
remnants, according to Eq.(\ref{eq:J}), is
\begin{equation}
\tau\sim\frac{x_i^3}{3g} t_{ch}\sim 5\times 10^{-10}{\rm sec}\ . \label{eq:U}
\end{equation}
The ``black hole epoch" thus ended in time for baryogenesis and other
subsequent epochs in the standard cosmology. As suggested in Ref.21, such a
post-inflation PBH evaporation provides an interesting mechanism for
reheating.

\section{Black Hole Remnants as Dark Matter}

This process also provides a natural way to create cold dark
matter. Although in our example $P(m)\sim 10^{-2}$, PBHs would
soon dominate the energy density by the time $t\sim
P(m)^{-2}t_h\sim 2\times 10^{-29}$s, because the original
relativistic particles would be diluted much faster than
non-relativistic PBHs. By the time $t\sim\tau$, all the initial BH
mass ($x_i$) had turned into radiation except one unit of $M_p$
preserved by each BHR. As BH evaporation rate rises sharply
towards the end, the universe at $t\sim \tau$ was dominated by the
BH evaporated radiation.

Roughly, $\Omega_{{\rm BHR},\tau}\sim 1/x_i$ and $\Omega_{\gamma,\tau}\sim 1$
at $t\sim \tau$, and since the universe resumed its standard evolution after
the black hole epoch ($t>\tau$), we find the density parameter for the BHR at
present to be
\begin{equation}
\Omega_{{\rm BHR},0}\sim
\Big(\frac{t_{eq}}{\tau}\Big)^{1/2}\Big(\frac{t_0}{t_{eq}}
\Big)^{2/3}\frac{1}{x_i}\Omega_{\gamma,0}\ , \label{eq:V}
\end{equation}
where $t_0\sim 4\times 10^{17}$s is the present time, and $t_{eq}$ is the
time when the density contributions from radiation and matter were equal. It
is clear from our construction that $(t_{eq}/\tau)^{1/2}\sim x_i$. So
$t_{eq}\sim 10^{12}$ sec, which is close to what the standard cosmology
assumes, and Eq.(\ref{eq:V}) is reduced to a simple and interesting
relationship:
\begin{equation}
\Omega_{{\rm BHR},0}\sim
\Big(\frac{t_0}{t_{eq}}\Big)^{2/3}\Omega_{\gamma,0}\sim
10^4\Omega_{\gamma,0}\ . \label{eq:W}
\end{equation}
In the present epoch, $\Omega_{\gamma,0}\sim 10^{-4}$. So we find
$\Omega_{{\rm BHR},0}\sim {\mathcal O}(1)$, about the right amount for dark
matter!

\section{Black Hole Epoch and Reheating Temperature}

As discussed above, shortly after PBHs were produced the density of the
universe was dominated by the BHs. Eventually the universe was reheated
through their continuous evaporation. To simplify the discussion we ignore BH
accretions of the radiation as well as BH mergers. Then under Hubble
expansion the effective reheating temperature at the end of the black hole
epoch, or $t\sim \tau$, can be expressed as
\begin{equation}
T_{r}(\tau[x_i])=\frac{1}{x_i-1}\int_1^{x_i}dx T_{\rm
GUP}(x)\frac{a(t[x])}{a(\tau[x_i])}\ , \label{eq:X}
\end{equation}
where $a(t)$ is the scale factor. Since $x_i \gg 1$, the evaporation only
became effective near the late times during this black hole epoch, when the
energy density was dominated by the BH radiation. As a further approximation
we assume radiation dominance throughout the BH epoch so that $a(t)\propto
t^{1/2}$. Expressing $t$ in terms of $x$ using Eq.(\ref{eq:I}), we find
\begin{equation}
T_{r}(\tau[x_i])\approx\frac{M_pc^2}{16\pi(x_i-1)}\Big[2\log(2x_i)-1\Big]+
{\mathcal O}\Big(\frac{1}{x_i^3}\Big)\ . \label{eq:Y}
\end{equation}
In our model $x_i\sim 4\times 10^{10}$. So $T_{r}(\tau[x_i])\sim 1.3\times
10^8$ GeV, which is sufficiently lower than the Planck and the GUT scales,
but higher than the baryogenesis scale.

\section{Acknowledgements}
I deeply appreciate my early collaborations and fruitful
discussions with Ronald J. Adler. I also thank S. Dimopoulos, A.
Green, A. Linde, M. Shmakova, and K. Thompson for helpful
discussions. This work is supported by the Department of Energy
under Contract No. DE-AC03-76SF00515.

\end{document}